\documentclass[prd,aps,floats,twocolumn,nofootinbib]{revtex4-1}
\usepackage{slashed}
\usepackage{mathtools}
\usepackage{amsfonts}
\usepackage{amssymb}
\usepackage{epsfig}
\usepackage{empheq}

\usepackage{bm}

\begin{document}
\newcommand*\widefbox[1]{\fbox{\hspace{2em}#1\hspace{2em}}}
\newcommand{\m}[1]{\mathcal{#1}}
\newcommand{\nn}{\nonumber}
\newcommand{\ph}{\phantom}
\newcommand{\eps}{\epsilon}
\newcommand{\be}{\begin{equation}}
\newcommand{\ee}{\end{equation}}
\newcommand{\bea}{\begin{eqnarray}}
\newcommand{\eea}{\end{eqnarray}}
\newtheorem{conj}{Conjecture}

\newcommand{\plk}{\mathfrak{h}}

%%%%%%%%%%%%%%%%%%%%%%%%%%%%%

%\title{wave function of the Universe with a variable cosmological constant}
\title{Space-time symmetry breaking on non-geodesic leaves and a new form of matter}
%[Global interactions, space-time] symmetry restoration and dark matter \\
%Dark matter due to symmetry restoration post-global interactions}
%Evolution and energy and momentum generation}
%without violation of local Lorentz invariance}

%Time as the chemical potential of fundamental constants}
%Generic evolving laws and energy creation}
%creation and annihilation}
%Varying speed of light and evolution in unimodular-like time}
\date{}

\author{Jo\~{a}o Magueijo}
\email{j.magueijo@imperial.ac.uk}
\affiliation{Theoretical Physics Group, The Blackett Laboratory, Imperial College, Prince Consort Rd., London, SW7 2BZ, United Kingdom}

\begin{abstract}
We examine the permanent damage caused by the historical breakdown of full diffeomorphism invariance induced by a  foliation. We focus on the case where the foliation is allowed to be non-geodesic after the interactions with the foliation switch off. Gravity and other forms of matter recover full diffeomorphism invariance only at the expense of introducing a new matter-like component, carrying the non-vanishing Hamiltonian (and momentum, as it turns out) left over from the violating past interactions. This matter form must be stress-free in the preferred frame; this is the only way a matter action can mimic the evolution of the leftover Hamiltonian (and momentum) driven by the Dirac hypersurface deformation algebra. Hence, if the preferred frame is 
non-geodesic, the equivalent matter component must have energy and a momentum current in this frame, but still no spatial stresses: an unusual form of ``matter''. It is equivalent to a fluid with anisotropic stress in some regimes, reducing to dust in others, or even displaying completely new features in extreme situations. Its stress energy tensor is conserved. We provide two examples based on accelerated frames: Rindler space-time and the canonical Schwarzchild frame. 
\end{abstract}

\maketitle

\section{Introduction}

This paper is a sequel to~\cite{geoCDM} where, on very general grounds, a relation was established between the enigma of dark matter and a possible primordial breakdown of 4D diffeomorphism invariance (whilst preserving 3D spatial diffeomorphism invariance), 
followed by symmetry restoration (for background to this work see~\cite{evol,BHevol,MachianCDM} with roots on~\cite{HL,shinji,NiaHL,Nia,Isham,brownkuchar,viqar,StueckelDM,Barvinsky,DBI} and~\cite{JoaoLetter,JoaoPaper,unimod,unimod1,UnimodLee1,alan,daughton,sorkin1,sorkin2,Bombelli,UnimodLee2,pad,pad1,lomb,vikman,vikman1,vikaxion}). It was shown that the degree of freedom associated with the smaller symmetry persists after symmetry restoration, carrying a memory of the integrated past effects of the symmetry breaking interactions. However, gravity and (normal) matter can be made to regain full symmetry and the correct number of degrees of freedom at the expense of introducing a new ``effective matter'' component, carrying the memory of the symmetry breaking phase. This component is equivalent to a pressureless fluid with 3 out of its 4 degrees of freedom frozen, pre-fixing its rest frame. A connection is thus made with cold dark matter, although differences exist which lead to new phenomenology, as explored in~\cite{geoCDM}.

Such symmetry breaking is effectuated by the introduction of a pre-fixed foliation, splitting space-time into spatial leaves and a {\it global} time label. While the symmetry breaking is active, the local Hamiltonian constraint (associated with the freedom to refoliate or to {\it locally} redefine time) is therefore lost, rendering the Hamiltonian locally non-zero. After symmetry restoration,  the non-vanishing value of the local Hamiltonian acts as a ``memory'' of the symmetry breaking phase. The question is then how to propagate this initial condition after symmetry restoration. In addressing this question, a central assumption made in~\cite{geoCDM} was that the preferred foliation is geodesic even after the full degrees of freedom of the metric are released (specifically, after the lapse function is allowed to be space-dependent in the variational calculus problem). We dubbed this assumption an ``equivalence principle for foliations'' and it sounds reasonable. But it is merely a possibility. This seemingly innocuous input has severe implications if dropped, as we demonstrate in this paper.

We recall that a geodesic foliation can be equivalently defined in terms of its normal $n_\mu$ (by requiring that it be geodesic), or in terms of its associated lapse function $N$ (by requiring that it be purely time dependent). Indeed the foliation's acceleration can be written: 
\begin{equation}\label{amueq}
    a_\mu =n^\nu\nabla_\nu n_\mu=h_\mu^{\;\nu} \nabla_\nu \ln N
\end{equation}
with $h_{\mu\nu}=g_{\mu\nu}+n_\mu n_\nu$ the  projector orthogonal to $n_\mu$ (see e.g.~\cite{ADMReview}). Its vanishing therefore implies
%\begin{equation}
   $ n_\mu=\partial_\mu T=(-N,0)$ for some scalar $T$, or $N=N(t)$.  
%\end{equation}
Hence, a non-geodesic foliation is one where the lapse function defined by the foliation is space dependent. In the language of Horava-Lifshitz theory~\cite{HL}, this dichotomy is referred to as a `projectable'' or ``non-projectable'' theory.

We will show in this paper that an on-shell spatially dependent lapse function introduces  violations of the local momentum constraint after symmetry restoration, even if the symmetry breaking phase only violated the local Hamiltonian constraint. This can be seen from the central equations (51) and (52) in~\cite{geoCDM}, re-derived and re-examined here (Eqs.~\eqref{dotHH} and \eqref{dotHHi}; specifically the first term on the right hand side of \eqref{dotHHi}). In addition, we show that in general, if we want to shift the blame of non-vanishing Hamiltonian and momentum to an effective matter component, then in the preferred foliation this matter component must be a stress-free. As proved in Section~\ref{Sec:nostress}, this is a pre-condition for any matter component to mimic the evolution of the Hamiltonian and momentum dictated by the Dirac hypersurface deformation algebra~\cite{Dirac,DiracCanadian,Thiemann,Bojo}.

Thus, we find that, in the case of a non-geodesic preferred foliation, the legacy matter component cannot have pressure or stresses, but given its non-vanishing momentum this frame cannot be its rest frame. A dust equivalent is therefore disqualified. Instead we find that the legacy effects of symmetry breaking are equivalent to a  matter component which has energy and momentum in the preferred foliation, but no stresses. In some regimes, this can be interpreted as anisotropic stresses in a rest frame that does not align with the preferred foliation. But perhaps this is best conceived simply as a new form of matter.  

%Significantly different phenomenology is therefore expected.   

\section{Set up}\label{setup}
Let us consider a theory with action $S_0$ which at some point in the past had its 4D diffeomorphism invariance downgraded to 3D diffeomorphisms on a foliation $\Sigma_t$ plus a global $t$ reparameterization invariance. For a concrete example of this, we can copy over the set up of global interactions at the start of~\cite{geoCDM} (where $S=S_0+S_{NL})$, specifically the separation of the Poisson bracket into the local variable terms and the global or non-local terms that break diffeomorphism invariance:
\begin{equation}
    \{.,.\}= \{.,.\}_L+ \{.,.\}_{NL}.
\end{equation}
Alternatively, we can ignore where past symmetry breaking came from and simply start from the end product, as we will presently see. 

We assume that $S_0$ has a metric which under an ADM split adapted to $\Sigma_t$ leads to lapse $N$, shift $N^i$, and induced or spatial metric (or projector) $h_{ij}$. The extended Hamiltonian density is:
\begin{equation}
    {\cal H}_E=N {\cal H} + N^i{\cal H}_i
\end{equation}
where ${\cal H}$ is the Hamiltonian density and ${\cal H}_i$ is the momentum density. Time derivatives of phase space variables are obtained from the PB with the total 
Hamiltonian $\mathbf{H}$: 
\begin{align}
     {\bf H}&=\int_{\Sigma_t} d^3 x\, {\cal H}_E(x).
\end{align}
We further assume that under the {\it local} Poisson bracket, $\cal H$ and ${\cal H}_i$ form the Dirac or hypersurface deformation algebra~\cite{Dirac,DiracCanadian,Thiemann,Bojo}, which we will use in its smeared form:
\begin{align}
    \{H_i(f^i), H_j(g^j)\}_L&= H_i([f,g]^i)\label{smearhihi}\\
    \{H_i(f^i), H(g)\}_L&= H(f^i\partial _ig )\label{smearhih0}\\
    \{ H(f), H(g)\}_L&= H_i(h^{ij}(f\partial_j g- g\partial_j f)) \label{smearh0h0}
\end{align}
where 
\begin{align}
    H(f)&=\int d^3 y\,  f(y) {\cal H}(y)\nn\\ H_i(f^i)&=\int d^3 y\,  f^i (y){\cal H}_i(y)\nn
\end{align}
and the test functions $f$ and $g$, and $f^i$  and $g^i$, can be any 3D scalar and vector, respectively. 
%Notice that the smeared $H(f)$ and $H_i(f^i)$ are 3D scalars and vectors (unlike the unsmeared ones, which are densities of weight 1). 

Given the relation between this algebra and diffeomorphisms, the statement that $S_0$ has 4D diffeomorphism invariance if global variables can be ignored is the statement that the theory satisfies the Dirac algebra under the local part of the Poisson bracket, but not under the full bracket. Hence $S_0$ could be as simple as plain General Relativity; but it could not be Horava-Lifshitz theory (studied in this context in~\cite{HLPaolo}).

If the theory had never had its 4D diffeomorphism invariance broken, then ${\cal H}(x)={\cal H}_i(x)=0$, following from $N$ and $N^i$ being Lagrange multipliers which are generically functions of space and time. By breaking the symmetry to 3D diffeomorphisms at some point, $N$ must be a function of time only (in general and not just on-shell), and so we lose the local constraint ${\cal H}(x)=0$, while preserving ${\cal H}_i(x)=0$ and a global constraint $\int dx\, {\cal H}(x)=0$.  Under the presence of global interactions respecting 3D diffeomorphism symmetry
(so that $\{ {\cal H}_i,\mathbf{H}\}=0$) we have~\cite{geoCDM}:
\begin{align}
      \dot {\cal H}&=\{ {\cal H},\mathbf{H}\}_{L} + \{ {\cal H},\mathbf{H}\}_{NL}\nn\\
       \dot {\cal H}_i&=\{ {\cal H}_i,\mathbf{H}\}_{L} + \{ {\cal H}_i,\mathbf{H}\}_{NL}=0.\nn
\end{align}
When the global interactions switch off (defined 
as $\{ {\cal H},\mathbf{H}\}_{NL}=0$) we are left with:
\begin{align}\label{initialcdns}
    {\cal H}(x,t_0)&\neq 0\nn\\
    {\cal H}_i(x,t_0)&= 0
\end{align}
which is the advertised ``end product'' of symmetry break. 

This end product can then be seen as initial conditions for an undriven evolution of ${\cal H}$ and ${\cal H}_i$ specified uniquely by the Dirac algebra, equations
\eqref{smearh0h0}-\eqref{smearhihi}. These imply:
\begin{align}
   \dot {\cal H}=\{{\cal H},\mathbf{H}\}_L&=\partial_i(N^i {\cal H})+\partial_i ({\cal H}^i N)+ {\cal H}^i \partial_i N \label{dotHH}\\
        \dot {\cal H}_i=\{{\cal H}_i,\mathbf{H}\}_L&= {\cal H}\, \partial_i N +D_j(N^j {\cal H}_i)  + {\cal H}_jD_i N^j\nn\\
        &= {\cal H}\, \partial_i N +\partial_j(N^j {\cal H}_i)  + {\cal H}_j\partial _i N^j,
        \label{dotHHi}
\end{align}
where ${\cal H}^i=h^{ij}{\cal H}_j$ 
and $D_i$ is the covariant derivative compatible with $h_{ij}$. To derive these equations we follow the convention spelled out in~\cite{Isham} whereby $\delta(x,y)$ is a bidensity, i.e.: a scalar as a function of the first argument, and a density as a function of the second\footnote{We thank Chris Isham for pearls of wisdom regarding this matter and others.}. 
We then set the first smearing functions to:
\begin{align}
    f(y)&=\delta(y,x)\nn\\
    f^i(y)&=\delta^i_j\delta(y,x)\nn
\end{align}
(so that these are indeed of weight zero, seen as test functions in $y$) and the second smearing function to the actual lapse scalar ($g(y)=N(y)$) and shift vector ($g^i=N^i(y)$). Given the density weights of the relevant objects, all $\partial_i$ in \eqref{dotHH} and \eqref{dotHHi} can be replaced by $D_i$ term by term, except for the last two terms in \eqref{dotHHi}, where this has to be done to the two terms together.

%As already suggested, we can ignore the framework of global interactions and simply assume that some symmetry breaking mechanism left us with \eqref{initialcdns} and \eqref{smearh0h0}-\eqref{smearhihi}, and so \eqref{dotHH} and \eqref{dotHHi}.

The fact that the Hamiltonian remains non-zero even after the source of symmetry breaking has been withdrawn might seem a permanent blemish. This is not the case. 
We can reinterpret the non-vanishing ${\cal H}$ (and ${\cal H}_i$, as we will see) as a new ``effective'' matter component with  contributions ${\cal H}_m$ and ${\cal H}^m_i$, so that:
\begin{align}
     \bar {\cal H}&={\cal H}(q(x),p(x);\alpha) +{\cal H}_m\approx 0\\
     \bar {\cal H}_i&={\cal H}_i(q(x),p(x);\alpha) +{\cal H}_i^m\approx 0.
\end{align}
These redefined quantities therefore satisfy local constraints. The redefined action $\bar S_0$ built from $\bar {\cal H}$ and $\bar {\cal H}_i$:
\begin{align}\label{barS0}
     \bar S_0=&S_0+S_m,
\end{align}
should therefore have a general lapse and shift functions:
\begin{align}
     N&=N(t,x)\label{Nofx}\\
      N^i&=N^i(t,x)
\end{align}
in order to enforce the newly found local constraints. Thus, full 4D diffeomorphism invariance is restored. The price to pay is the new form of matter $S_m$. 

%In this way, when ${\cal H}_m$ and ${\cal H}^m_i$ are negligible, we recover full 4D diff invariance, and  when it is not, the new form of matter is the scapegoat for all the blemishes left by past symmetry breaking. 

%[MUST IMPROVE THE BACKGROUND TO THIS]. This is the action we started from [???], but with $N=N(t,x)$, that is with full [COPY OVER HERE THE BIT ON [Add here the bit of dofs of gravity in vacuum always]]

For this to work, this type of matter must have an evolution with  ${\cal H}_m$ and ${\cal H}^m_i$ satisfying:
\begin{align}
   \dot {\cal H}_m&=\partial_i(N^i {\cal H})+\partial_i ({\cal H}^i N)+ {\cal H}^i \partial_i N \label{dotHHm}\\
        \dot {\cal H}^m_i&= {\cal H}_m\, \partial_i N +D_j(N^j {\cal H}^m_i)  + {\cal H}^m_j D_i N^j.\label{dotHHim}
\end{align}
The fork highlighted in the introduction is now crucial: whether $\Sigma_t$ is geodesic or not. Once the full degrees of freedom of the metric are released (cf. \eqref{Nofx}) we have to decide whether on-shell for $\Sigma_t$ we still impose $N=N(t)$, corresponding to a  geodesic $\Sigma_t$, or not. The former case implies that if initially ${\cal H}_i=0$ (cf.~\eqref{initialcdns}), then this is preserved by the evolution. We then have a dust fluid with a rest frame coinciding with $\Sigma_t$, as studied in~\cite{geoCDM}. If $\Sigma_t$ is non-geodesic, then $\partial_i N\neq 0$ on-shell, so that a non-vanishing ${\cal H}$ will always drive a non-vanishing ${\cal H}_i\neq 0$ on $\Sigma_t$ (cf. the first term in the right hand side of \eqref{dotHHi} or \eqref{dotHHim}). This is the more general case to be studied in this paper.

\section{``No-stress condition'' for a matter representation}\label{Sec:nostress}

We first derive generic conditions upon this type of effective matter for it to serve its purpose. Its evolution, given by Eqs.~\eqref{dotHHm} and \eqref{dotHHim}, must mimic Eqs. \eqref{dotHH} and \eqref{dotHHi}. However, the latter two equations fundamentally concern ${\cal H}$ and ${\cal H}_i$ associated with the {\it total} Hamiltonian and momentum, given by gravity plus ``real'' matter, with their evolution resulting from Poisson brackets involving all pairs of local variables, including ``real'' matter and gravity. 
This is what we want to mimic with a new ``fictitious'' matter component.  

However, \eqref{dotHHm} and \eqref{dotHHim} are generally {\it not} the equations of motion of the Hamiltonian and momentum of a matter component.
%even if it only interacts with gravity and not with other forms of matter. 
For a matter component which only 
couples to the metric (i.e. $N$ and $N^i$ via the usual constraints and $h_{ij}$, but not its conjugate $\Pi^{ij}$), we have:
\begin{align}\label{correct1}
    \dot {\cal H}_m&=
      \{{\cal H}_m,\bar {\mathbf{H}}\}_L=\{{\cal H}_m,{\mathbf{H}}_m 
    \}_{q_m,p_m} +\frac{\delta {\cal H}_m}{\delta h_{ij}}\dot h_{ij}\\
      \dot {\cal H}_i^m&=
      \{{\cal H}_i^m,\bar {\mathbf{H}}\}_L=\{{\cal H}_i^m,{\mathbf{H}}_m 
    \}_{q_m,p_m} +\frac{\delta {\cal H}_i^m}{\delta h_{ij}}\dot h_{ij}
    %\\***
    %\{{\cal H}_m,{\mathbf{H}}\}_{h_{ij},\Pi^{ij}}\nn
\end{align}
where $(q_m,p_m)$ are the matter component degrees of freedom, and the last term results from the gravity variables in the PB.
%\footnote{We obviously also assumed that  gravity and other matter do not depend on $(q_m,p_m)$.}. 

In addition, a generic form of matter does not separately satisfy the Dirac algebra \eqref{smearhihi}-\eqref{smearh0h0}, required for mimicking \eqref{dotHH} and \eqref{dotHHi}\footnote{This matter has been discussed in the literature, eg.\cite{brownkuchar} Sec.~IVA.}. The algebra is valid for the {\it total}  ${\cal H}$ and ${\cal H}_i$ as well as for their gravity contributions. Hence cross terms between gravity and each matter component spoil the algebra for each matter component. These cross-terms are zero for the PB \eqref{smearhihi} and \eqref{smearh0h0} under minimal assumptions; specifically that the gravity Hamiltonian does not contain derivatives of $\Pi^{ij}$ (as is the case of GR), or that the {\it matter} diffeomorphism constraint does not contain the metric
\begin{equation}\label{nostressHi}
    \frac{\delta {\cal H}_i^m}{\delta h_{ij}}=0
\end{equation}
as is the case with a dust, scalar fields, etc\footnote{Other assumptions could be made, for example with spinorial matter, but they will not be explored here. We thank Stephon Alexander for this remark.}. However, one of the cross terms is inevitably non-zero for the PB involving the momentum and the Hamiltonian,  \eqref{smearhih0}, specifically the term resulting in the correction:
\begin{align}
     \{H^m_i(N^i), H_m(N)\}&= H_m(N^i\partial _iN)- \{H^G_i(N^i), H_m(N)\}.\nn
\end{align}
Using the fact that $H^G_i(N^i)$ generates spatial diffeomorphisms we have\footnote{It can be checked that this identity is valid for concrete matter sources, for example a dust fluid or a scalar field.}:
\begin{align}
      \{H^m_i(N^i), H_m(N)\}&= H_m(N^i\partial _iN) \nn\\
     & +\int dx\, \frac{\delta {\cal H}_m}{\delta h_{ij}}N(D_i N_j + D_j N_i).\nn
\end{align}
Plugging this extra term into  \eqref{correct1} we get 
\begin{align}
   \dot {\cal H}_m&=[{\rm required}]-2N  \frac{\delta {\cal H}_m}{\delta h_{ij}} K_{ij}
   %\partial_i(N^i {\cal H}_m)+\partial_i (h^{ij}{\cal H}^m_j N)+ h^{ij}{\cal H}^m_i \partial_j N \label{dotHHm}\\
%        \dot {\cal H}^m_i&= ...+
        %{\cal H}_m\, \partial_i N +\partial_j(N^j {\cal H}^m_i)  + {\cal H}^m_j\partial_i N^j\label{dotHHim}
\end{align}
where ``required'' stands for the right hand side of \eqref{dotHHm} and $K_ij$ is the extrinsic curvature. %and \eqref{dotHHim}.

The upshot is that 
we need vanishing Eulerian spatial stresses, including pressure, in the preferred frame $\Sigma_t$:
\begin{equation}\label{nostress}
    \frac{\delta {\cal H}_m}{\delta h_{ij}}=0
\end{equation}
for the effective fluid to reproduce \eqref{dotHH}. It can be checked that reproducing \eqref{dotHHi} does not include new conditions (it requires the already assumed innocuous \eqref{nostressHi}, as well as \eqref{nostress} and its spatial derivatives).  Hence the central condition for a matter component to represent the evolution of a legacy violation of the Hamiltonian and momentum constraint is that the preferred frame $\Sigma_t$ must be the ``no-stress frame'' for this fluid (and obviously, that such frame exists for the putative matter component). 

For a geodesic $\Sigma_t$, where the momentum constraint vanishes (on $\Sigma_t$), this condition implies a dust effective fluid (no pressure, no anisotropic stresses in the preferred frame, identified with the dust rest frame). We have generalized the no-stress/pressure condition for cases where $\Sigma_t$ might not be geodesic and ${\cal H}_i^m\neq 0$.

\section{Properties of the effective matter component}\label{Sec:set}
We now prove that 
the condition derived in the previous Section rules out a perfect fluid if ${\cal H}^m_i\neq 0$ on $\Sigma_t$ (which is necessary if $\Sigma_t$ is non-geodesic, as we saw).  
The stress-energy tensor on $\Sigma_t$ for a matter component satisfying \eqref{nostress} is given by:
\begin{align}
    T^m_{00}&=\frac{-2}{N\sqrt{h}}\frac{\delta S_m}{\delta g^{00}}=
   \frac{N}{\sqrt h}(N  {\cal H}_m +2N^i {\cal H}^m_i) ,
   \nn\\
    T^m_{0i}&=\frac{-2}{N\sqrt{h}}\frac{\delta S_m}{\delta g^{0i}}=\frac{N}{\sqrt{h}}{\cal H}^m_i,\nn\\
    T^m_{ij}&=0,\label{Tmn}
\end{align}
(with $h=\det (h_{ij})$), 
where care must be taken to keep other components of the metric constant when varying each one of them. Hence, we can write:
\begin{align}\label{nostressT}
      T^m_{\mu\nu}&=\rho n_\mu n_\nu-2\Pi^\alpha n_{(\mu} h_{\nu)\alpha}\nn\\
    &=
    \rho n_\mu n_\nu-( n_\mu\Pi_\nu +   n_\nu\Pi_\mu) 
\end{align}
with
\begin{align}
    \rho&=n_\mu n_\nu T_m^{\mu\nu}=\frac{{\cal H}_m }{\sqrt{h}}\nn\\
    \Pi^\mu&= h^{\mu\alpha}n^\nu T^m_{\alpha\nu}={\left(0,\frac{{\cal H}^i_m}{\sqrt{h}}\right)}\label{nostressT1}
\end{align}
where, we recall, $n_\mu =(-N,0)$ is the normal 
and $h_{\mu\nu}=g_{\mu\nu}+n_\mu n_\nu$ the projector associated with $\Sigma_t$, 
and where we used:
\begin{align}
     {\cal H}_m&= \sqrt{h}n^\mu n^\nu T^m_{\mu\nu}\label{HinT}\\
     {\cal H}^m_i&=
     \sqrt{h}n^\mu h^\nu_{\; i} T^m_{\mu\nu}.\label{HiinT}
\end{align}
%[ALSO
%\begin{equation}
%     T^m_{\mu\nu}=\rho n_\mu n_\nu-(n_\mu \Pi_\nu + n_\nu \Pi_\mu)
%\end{equation}
%Thus (on $\Sigma_t$) there IS no term proportional to two spatial projectors, the only other possibility. OR:
If ${\cal H}_i^m=0$ on a geodesic preferred $\Sigma_t$, then $\Pi^\mu=0$, and we recover a dust fluid in its rest frame, as in~\cite{geoCDM}. Otherwise ${\cal H}^m_i\neq 0$ on $\Sigma_t$ and so $\Pi^\mu\neq 0$, implying a type of matter with an Eulerian energy current (or momentum density) but no Eulerian stresses. This property is rather peculiar: energy and momentum but no spatial stresses in one frame.

\subsection{Compare and contrast with dust}

This is in stark contrast with dust as seen away from its rest frame, for which Eulerian spatial stresses arise in tandem with a momentum current (see for example~\cite{brown}, or Section VII of~\cite{geoCDM}). Recall that for dust with a
rest frame $u$ generically different from $n$ we have: 
\be\label{setdust}
T^d_{\mu\nu}=\rho_d u_\mu u_\nu
\ee
where 
\begin{align}
    u_\mu&=(N^i u_i - N \gamma, u_i)\\
     u^\mu&=\left(\frac{\gamma}{N}, u^i-\frac{N^i\gamma}{N}\right)
\end{align}
with $u^i\equiv h^{ij} u_j$ and ``gamma factor''~\cite{brown} given by: 
\begin{equation}\label{gamma}
    -n^\mu u_\mu =\sqrt{1+h^{ij} u_i u_j}\equiv \gamma(u_i,h^{ij}).
\end{equation}
Setting:
\be
u^\mu=\gamma n^\mu +w^\mu
\ee
with $w^\mu=h^\mu_{\, \nu} u^\nu$ we find:
\begin{align}
    w^\mu&=(0,u^i), \nn\\
    w_\mu & =(N^i u_i,u_i)\label{wcoords}
\end{align}
as well as the useful identity
\begin{equation}
    \gamma^2-1=u_i u^i =w_\mu w^\mu.
\end{equation}
Expanding \eqref{setdust}:
\begin{equation}
    T^d_{\mu\nu}=\rho_d\gamma^2 n_\mu n_\nu + \rho_d \gamma (n_\mu w_\nu + n_\nu w_\mu) + \rho_d w_\mu w_\nu\nn 
\end{equation}
we do get the correct two terms in \eqref{nostressT}, with:
\begin{align}
    \rho&=\rho_d\gamma^2\nn\\
    \Pi_\mu&=-\rho_d\gamma w_\mu\label{DustTrans}
\end{align}
but also have an additional ``stress'' term, proportional to two projectors. The latter cannot exist in our putative fluid, as we proved in Section~\ref{Sec:nostress} (it would violate the condition \eqref{nostress}).

{\it In some cases} (the exceptions to be detailed presently)  we can think of it as dust with a correction:
\begin{equation}\label{setmdust}
    T^m_{\mu\nu}=T^d_{\mu\nu}-\frac{\Pi_\mu\Pi_\nu}{\rho}. 
\end{equation}
This is equivalent to a fluid with pressure:
\begin{equation}
    p=-\frac{1}{3} \rho_d (\gamma^2-1)
\end{equation}
and anisotropic stress:
\begin{equation}
    \Sigma_{\mu\nu}=-\frac{\Pi_\mu\Pi_\nu}{\rho}-ph_{\mu\nu}.
    %=-\rho_d\left[w_\muw_\nu  ETCETC TBD\right].
\end{equation}
That is, the fluid has a negative pressure aligned with the direction of ${\cal H}_i$, and zero pressure in the other directions. This anisotropic stress precludes a perfect fluid.  
For dust, with ${\cal H}_i=0$, the no-stress frame coincides with the rest frame, so this is the only case where we have a perfect fluid. 

\subsection{Recovery of CDM in the Newtonian limit}\label{Sec:lin}
Nonetheless in the Newtonian limit our fluid is indistinguishable from CDM: in the weak field non-relativistic limit the correction term in \eqref{setmdust} is negligible, because it is quadratic in the velocity $u^i$. 
%Hence in the Newtonian limit our fluid is indistinguishable from cold dark matter. 
This can be formalized by assuming that the preferred frame is the linearizing frame (which is non-geodesic), with approximate metric in isotropic coordinates given by:
\begin{equation}
    ds^2=-(1+\Phi_N) dt^2+(1-\Phi_N)(dx^2+dy^2+dz^2).
\end{equation}
Dropping the last term in \eqref{setmdust}, using $\gamma\approx 1$ and $\rho\approx \rho_d$, and projecting stress-energy conservation onto and orthogonal to $u^\mu$ we get: 
\begin{align}
    \dot \rho+\rho\partial_iu^i&=0\\
    \dot u^i &=-\partial_i\Phi_N 
\end{align}
where dots refer to convective (or Lagrangian) derivatives (i.e. $\partial/\partial t + v^i\partial_i$). These are the equations of motion of dust in the Newtonian limit, proving our claim. Indeed, since this fluid has all of its 4 degrees of freedom released, it does not even have the subtle distinguishing features described in~\cite{geoCDM} in the case of a geodesic $\Sigma_t$. 

\subsection{Breakdown of the dust/fluid analogy}
The analogy with a fluid (near-dust, or even a fluid with anisotropic stresses) breaks down completely if:
\begin{equation}\label{breakdown}
    h^{ij}{\cal H}^m_i{\cal H}^m_j\ge {\cal H}_m^2.
\end{equation}
This can happen, for example, as a result of the initial conditions left over by the non-local interactions, if these do not respect the momentum constraint.
%[CUT or due to the free evolution (as we will see)]. Indeed,
Eqs.~\eqref{nostressT1}, \eqref{wcoords} and \eqref{DustTrans} imply:
\begin{align}
    \frac{{\cal H}_m }{\sqrt{h}}&= \rho=\rho_d\gamma^2\\\
    \frac{{\cal H}_i^m}{\sqrt{h}}&=-\rho_d\gamma u_i=-\frac{\rho}{\gamma}u_i.
\end{align}
%[we could use $h_S/h=\gamma^2$]. 
From these (just as in~\cite{Bojo}, example 3.25) we can define the ``velocity'' as:
\begin{equation}\label{V}
    V_i=-\frac{{\cal H}_i}{\cal H}=\frac{u_i}{\gamma}= \frac{u_i}
    {\sqrt{1+u^2}}
    %{\sqrt{1+u_iu^i}}
\end{equation}
giving the gamma factor its more recognizable form:
\begin{equation}
    \gamma=\frac{1}{\sqrt{1-V^2}}
    %{\sqrt{1-V_i V^i}},
\end{equation}
and implying the reciprocal relation:
\begin{equation}
   u^2=\frac{V^2}{1-V^2}=V^2\gamma^2 
\end{equation}
(with $u^2$ and $V^2$ computed with $h_{ij}$). 
This shows that the analogy with an anisotropic fluid breaks down under \eqref{breakdown}. Then $V\ge 1$, implying an imaginary $u^i$ and $w_\mu$ and a negative $\rho_d$. the fluid analogy does break down and one is better off reverting to form  \eqref{nostress} in this regime.

\section{Stress-energy tensor conservation}\label{Sec:setcons}
We finally prove a very important property of the effective fluid: its stress-energy tensor is conserved. Indeed we will show that evolution equations for the Hamiltonian and momentum constraint, \eqref{dotHHm} and \eqref{dotHHim}, are equivalent to conservation along and orthogonal to $n^\mu$:
\begin{widetext}
\begin{align}
 n_\mu\nabla_\nu T_m^{\mu\nu}=0&\iff   \dot {\cal H}_m=\partial_i(N^i {\cal H}_m)+\partial_i ({\cal H}_m^i N)+ {\cal H}_m^i \partial_i N , \label{equiv1}\\
  h_{\alpha\mu}\nabla_\nu T_m^{\mu\nu}=0&\iff   
    \dot {\cal H}^m_i={\cal H}_m\, \partial_i N +\partial_j(N^j {\cal H}^m_i)  + {\cal H}^m_j\partial _i N^j, \label{equiv2}
\end{align}
\end{widetext}
supplying an interesting physical interpretation of the Dirac algebra and the evolution of the constraints. They are the Hamiltonian expression of stress-energy conservation (and ultimately of the generalized Bianchi identities, as we will discuss in~\cite{CDMLIV}). 

To prove this we will need the decomposition \eqref{nostressT} with \eqref{nostressT1}, always bearing in mind that $n_\mu\Pi^\mu=0$, $h^\mu_\alpha\Pi^\alpha=\Pi^\mu$. In addition to \eqref{nostressT1}, the components 
\begin{equation}
    \Pi_\mu =\frac{1}{\sqrt{h}}(N^k{\cal H}^m_k,{\cal H}^m_i)
\end{equation}
will also be useful, as well as the contraction:
\begin{equation}
    h_\mu^{\; \beta} T^m_{\beta\nu}=-\Pi_\mu n_\nu.
\end{equation}
The other possible single contraction of the stress-energy tensor is:
\begin{equation}
    P_\mu=T^m_{\mu\nu}n^\nu=\frac{1}{\sqrt{h}}(N{\cal H}_m +N^k{\cal H}^m_k,{\cal H}^m_i)\nn
\end{equation}
or   
\begin{equation}
    P^\mu=\frac{1}{{\sqrt{-g}}}(-{\cal H}_m, N^i{\cal H}_m+N{\cal H}_m^i)\label{pupmu}
\end{equation}
so that $n^\mu P_\mu=\rho$ and $h_\mu^{\; \nu} P _\nu=\Pi_\mu$. 
We can think of $\rho$ as the energy density (the scalarized version of the Hamiltonian),  $\Pi^\mu$ as the momentum density (a covariantized version of the Momentum constraint), and $P^\mu$ as the energy-momentum 4-vector, conflating the previous two. 

\subsection{Proof of the first equivalence}
We first prove equivalence \eqref{equiv1}. 
We start by writing:
\begin{align}
    n_\nu\nabla_\mu T_m^{\mu\nu}\nn &= \nabla_\mu (n_\nu T_m^{\mu\nu}) -T_m^{\mu\nu} \nabla_\mu n_\nu\nn\\
    &=\nabla_\mu P^\mu -T_m^{\mu\nu} \nabla_\mu n_\nu,\nn 
\end{align}
and since $K_{\mu\nu}=-(\nabla_\mu n_\nu+n_\mu a_\nu)$, $h^\mu_\alpha h^\nu_\beta T^m_{\mu\nu}=0$ and $h^\mu_\alpha a_\mu=a_\alpha$, we can write the last term as: 
\begin{align}
    -T_m^{\mu\nu} \nabla_\mu n_\nu=T_m^{\mu\nu}(n_\mu a_\nu + K_{\mu\nu})=T_m^{\mu\nu}n_\mu a_\nu=\Pi^\mu a_\mu\nn
\end{align}
resulting in the tensorial identity:
\begin{align}
    n_\nu\nabla_\mu T_m^{\mu\nu}
    &=\nabla_\mu P^\mu + \Pi^\mu a_\mu .\label{tensorial1}
\end{align}
Introducing components (as in \eqref{pupmu}) converts the first term in:
\begin{align}
     \nabla_\mu P^\mu&=\frac{1}{\sqrt{-g}}\partial_\mu ({\sqrt{-g}} P^\mu)\nn\\
     &= \frac{1}{\sqrt{-g}}(-\dot {\cal H}_m+\partial_i(N^i {\cal H}_m)+\partial_i (h^{ij}{\cal H}^m_j N)),\nn 
\end{align}
and \eqref{amueq} and \eqref{nostressT1} convert the second in:
\begin{align}
    \Pi^\mu a_\mu=\frac{1}{\sqrt{-g}} {\cal H}_m^i \partial_i N .\nn
\end{align}
Substituting in \eqref{tensorial1} therefore proves the equivalence \eqref{equiv1}.

\subsection{Proof of the second equivalence}
We next prove equivalence \eqref{equiv2}. First we rewrite:
%the LHS of the equivalence as:
\begin{align}
      h_{\alpha\mu}\nabla_\nu T_m^{\mu\nu}&=
      \nabla_\nu ( h_{\alpha\mu} T_m^{\mu\nu})- (\nabla_\nu h_{\alpha\mu}) T_m^{\mu\nu}
      \nn\\
      &=-\nabla_\mu (n^\mu \Pi_\alpha) +\rho a_\alpha +\Pi^\nu K_{\nu\alpha}.\nn
\end{align}
The projector is idempotent, so we will have achieved our task by proving that:
\begin{align}\label{temp1}
    h_\alpha^\beta\nabla_\mu (n^\mu \Pi_\beta) = \rho a_\alpha +\Pi^\nu K_{\nu\alpha}
\end{align}
becomes proportional to:
\begin{align}\label{evolHi}
    \dot {\cal H}^m_i-\partial_j(N^j {\cal H}^m_i) ={\cal H}_m\, \partial_i N  + {\cal H}^m_j\partial _i N^j
\end{align}
term by term except for connection terms that cancel out. 

In order to do this, 
first we note that \eqref{temp1} is a transverse identity, so its full content can be found by setting $\alpha=i$. Then we write:
\begin{align}
\nabla_\mu (n^\mu \Pi_i) &=\frac{1}{\sqrt{-g }}
({\sqrt{-g }}n^\mu \Pi_i)_{,\mu }-\Gamma^\beta_{\mu i}n^\mu \Pi_\beta \nn\\
 &=\frac{1}{\sqrt{-g }}( \dot {\cal H}^m_i-\partial_j(N^j {\cal H}^m_i)) -\Gamma^\beta_{\mu i}n^\mu \Pi_\beta,
 \end{align}
 and
 \begin{align}
     \rho a_i&=\frac{1}{\sqrt{-g}}{\cal H}_m\partial_i N.
 \end{align}
The last term in \eqref{temp1} can be manipulated as:
\begin{align}
    \Pi^\nu K_{\nu i}&= 
    -\Pi^\nu h^\alpha_\nu \nabla_i n_\alpha 
    =-\Pi^\alpha \nabla_i n_\alpha =- \Pi_\nu \nabla_i n^\nu \nn
\end{align}
since $K_{\mu\nu}$ is symmetric and can be obtained from a single projector applied to $\nabla_\mu n_\nu$, instead of the two that appear in its definition.  This can then be expanded as:
\begin{align}
    - \Pi_\nu \nabla_i n^\nu 
    &=-\Pi_{\nu}(\partial_in^\nu +\Gamma_{i\mu}^\nu n^\mu)\nn\\
    &=-\frac{N^k{\cal H}^m_k}{\sqrt{h}}\partial_i\frac{1}{N}+ \frac{{\cal H}^m_k}{\sqrt{h}}\partial_i\frac{N^k}{N} 
    - \Gamma_{\mu i}^\beta n^\mu\Pi_\beta\nn\\
    &=\frac{{\cal H}^m_k}{\sqrt{-g}}\partial_i N^k 
     - \Gamma_{\mu i}^\beta n^\mu\Pi_\beta.
\end{align}
Collecting all the terms we find the promised cancellation of the cumbersome connection terms, leaving us with a term by term correspondence between \eqref{temp1} and \eqref{evolHi} (up to a common factor). 

\section{Examples}\label{examples}
We close by providing examples of how the general results derived in this paper would be realized in concrete examples of non-geodesic foliations. We focus on the simple case where the effective matter acts as a test body, generalizing the exercise performed in~\cite{geoCDM} for geodesic frames.

\subsection{Rindler space-time}
If we take for local preferred frame Minkowski space (the background to the linearizing exercise in Section~\ref{Sec:lin}) then the evolution is trivial:  $\dot{\cal H}=\dot{\cal H}_i=0$, so that any non-vanishing initial conditions are simply preserved. But what if the preferred frame were accelerated with respect to this frame, or indeed the asymptotic cosmological frame? 

As a model for such an eventuality we take the Rindler frame:
\begin{equation}
    ds^2=-(\alpha x)^2 dt^2 + dx^2 + dy^2 +dz^2
\end{equation}
(say, the $x>|t|$ wedge), 
so that $N=\alpha x$, $N^i=0$, $h_{ij}={\rm diag} (1,1,1)$. The evolution equations \eqref{dotHHm} and \eqref{dotHHim} then become:
\begin{align}
    \dot {\cal H}&=\alpha x\partial_x{\cal H}_x + 2\alpha {\cal H}_x \\
    \dot {\cal H}_x&=\alpha {\cal H}\\
     \dot {\cal H}_y&=0\\
     \dot {\cal H}_z&=0.
\end{align}
We assume ${\cal H}_i=0$ at the initial time $t=0$ when free evolution started; then only ${\cal H}_x$ can become non-vanishing at $t>0$. We can derive a second order equation for ${\cal H}$:
\begin{equation}
    \ddot {\cal H}=\alpha^2 x\partial_x {\cal H}+2\alpha^2 {\cal H}
\end{equation}
showing that there is an instability. For example, an initially homogeneous ${\cal H}(t_0,x)={\cal H}_0$ evolves as:
\begin{align}
    {\cal H}&={\cal H}_0\exp{\left(\sqrt{2}\alpha t\right)}.  
\end{align}

We can map these results into the fluid analogy of Section~\ref{Sec:set}. Eq.~\eqref{Tmn} leads to stress energy tensor:
\begin{align}
    T_{00}&=(\alpha x)^2{\cal H}\nn\\
    T_{0x}&=\alpha x{\cal H}_x.
\end{align}
It can be checked that this is conserved (as it should be; cf. Section~\ref{Sec:setcons}). Comparing with \eqref{nostressT} and \eqref{nostressT1} we find:
\begin{align}
     \rho&={\cal H}={\cal H}_0\exp{\left(\sqrt{2}\alpha t\right)}\\
     \Pi_x&={\cal H}_x=\frac{{\cal H}_0}{\sqrt{2}}\left(\exp{\left(\sqrt{2}\alpha t\right)}-1\right).
\end{align}
Using \eqref{V}, we see that such fluid would tend to a uniform velocity in the Rindler frame:
\begin{equation}
    V_x=-\frac{1}{\sqrt{2}}\left(1-\exp{\left(-\sqrt{2}\alpha t\right)}\right)\rightarrow -\frac{1}{\sqrt{2}}
\end{equation}
that is $\gamma\rightarrow \sqrt{2}$. The analogy with an anisotropic fluid therefore works, with ``rest'' energy density
\begin{equation}
    \rho_d=2\rho=2{\cal H}=2{\cal H}_0\exp{\left(\sqrt{2}\alpha t\right)} 
\end{equation}
velocity 4-vector:
\begin{equation}
    u^\mu =\left(\frac{2}{\alpha x},1,0,0\right)
\end{equation}
and anisotropic pressure along the $x$ direction:
\begin{equation}
    p_x=-2\rho=-\rho_d
\end{equation}
with $p_y=p_z=0$.

%This is very interesting: even in the case of plain General Relativity points to an instability in the first class constraints if expressed in a non-geodesic foliation. Physically, on the other hand, it is possible that a this feature might assist in a self-procreating? model of the Early Universe. ????

\subsection{The Schwarzchild frame}
As another example, we study the passive gravity of the effective fluid around the outside of the horizon of a black hole foliated with Schwarzchild leaves, the counterpart of the example given in \cite{geoCDM} (with the geodesic foliation provided by the Lemaitre-Tolman-Bondi frame). 
Then:
\begin{align}
    N^2&=1-\frac{2m}{r}\nn\\
    N^i&=0\nn\\
    h_{ij}&={\rm diag}(1/N^2,r^2,r^2\sin^2\theta)
\end{align}
so that \eqref{dotHH} and \eqref{dotHHi} reduce to:
\begin{align}
       \dot {\cal H}(t,r)&=
       %\partial_r (N^3{\cal H}_r )+ N^2{\cal H}_r \partial_r N 
       \frac{1}{N} \partial_r(N^4 {\cal H}_r)\nn\\
        \dot {\cal H}_r(t,r)&= {\cal H}\, \partial_r N
\end{align}
(where we have dropped the label $m$). These can be combined into:
\begin{align}\label{ddotHSch}
    \ddot {\cal H}=\frac{1}{N}\partial_r\left( N^4 {\cal H}\partial_r N\right).
    %\partial_r\left( N {\cal H}\right).
\end{align}
It can be checked that the associated stress-energy tensor:
\begin{align}
    T_{00}&=\frac{N^3}{r^2\sin\theta}{\cal H}\nn\\
    T_{0r}&=\frac{N^2}{r^2\sin\theta}{\cal H}_r
\end{align}
is again conserved (as proved in general). 

In the Newtonian limit this is just dust, so there are no novelties. The differences arise if the ``velocity'' $V_r$ becomes comparable or larger than 1. The most extreme location for this to happen in near the horizon, $r\approx 2m$, and regime we now explore. Then, \eqref{ddotHSch} reduces to:
\begin{equation}
      \ddot {\cal H}\approx \frac{3}{16m^2}{\cal H} + \frac{N^2}{4m}\partial_r{\cal H}
\end{equation}
if we allow any $\epsilon=r-2m$ dependence in $\cal H$. For a Black hole formed around $t=0$ with a uniform leftover Hamiltonian as initial condition, we have for a possible growing mode:
\begin{align}\label{HamBH}
    {\cal H}&={\cal H}_0\exp{\left(\frac{\sqrt{3}}{4m}t\right)}.
\end{align}
Thus the momentum constraint diverges at the horizon, since  $\dot {\cal H}_r(t,r)= {\cal H}\, \partial_r N\approx {\cal H}/(4m N)$, but this leads to
\begin{equation}
    V_r=-\frac{{\cal H}_r}{\cal H}=-\frac{1}{N\sqrt{3}}\implies V^2=\frac{1}{3}
\end{equation}
or $\gamma^2=3/2$. An anisotropic fluid analogy is therefore possible. We can read off:
\begin{align}
    \rho&=\frac{N{\cal H}_0}{16m^2\sin\theta}\exp{\left(\frac{\sqrt{3}}{4m}t\right)}\label{rhoBH}\\
    \Pi_r&=\frac{1}{\sqrt{3}}\frac{N{\cal H}_0}{16m^2\sin\theta}\exp{\left(\frac{\sqrt{3}}{4m}t\right)}
\end{align}
mapping into:
\begin{align}
    \rho_d&=\frac{2}{3}\rho\\
   u^\mu&=\left(-\sqrt{\frac{3}{2}}N, -\frac{N}{\sqrt{2}}\right).
\end{align}
On each leave (with constant $t$) this solution shows a smooth transition to $\rho=0$ at the boundary of the foliation $r\rightarrow 2m$. Any observer, however, would see all the leaves up to $t\rightarrow\infty$ as $r=2m$ is approached. The exponential in \eqref{rhoBH} would therefore win over the power law coming from the determinant, and a soft singularity would be experienced.

\section{Conclusions}
In this paper we continued the work of~\cite{geoCDM} to investigate the effects of past 4D symmetry breaking followed by restoration, with the significant difference that the foliation breaking the symmetry is allowed to be non-geodesic. %even if only after symmetry restoration. 
In such a case, the regained Dirac algebra implies that the local momentum constraint must be violated along with the Hamiltonian constraint, after the sources of symmetry violation are withdrawn (see Section~\ref{setup}). It is possible to transfer these violations to an effective matter fluid, so that the theory recovers full symmetry, but this requires that the fluid must not have spatial stresses or pressure. Only then does the fluid evolution mimic the evolution predicted by the Dirac algebra, as discussed in Section~\ref{Sec:nostress}. This no-stress condition implies a dust fluid in the geodesic case studied in~\cite{geoCDM} (for which the momentum constraint is enforceable, so rendering $\Sigma_t$ the dust rest frame, for which there are no stresses). The implications for a non-geodesic frame are more dramatic, and point towards a new type of (dark) matter. 

As shown in Section~\ref{Sec:set}, on a non-geodesic $\Sigma_t$ the no-stress condition implies a type of effective matter which in one frame (the preferred $\Sigma_t$) would be observed as possessing energy density and Eulerian energy current (or momentum density) but no Eulerian stresses. This property is rather peculiar: energy and momentum, but no spatial stresses in one frame. It cannot be associated with dust as seen away from its rest frame, for which stresses always arise together with momentum density. Indeed it precludes any type of perfect fluid. It can be interpreted as a fluid with anisotropic stresses in some regimes, but the analogy breaks down in others. If the momentum is much larger than the Hamiltonian the ``effective velocity'' would be superluminal. This is an effective fluid and not a real matter component, so just as one can argue that negative energy is not a serious impediment~\cite{shinji},  faster than light speeds do not need to have signal communication implications. Alternatively one might simply view this extreme regime and a break down of the anisotropic fluid interpretation. In spite of these oddities, we can also define a ``Newtonian'' regime where this form of matter would be indistinguishable from dust. Indeed this is how our results reduce to those in~\cite{geoCDM} in some appropriate limit, but this is far from generic. 

An important consistency check is conservation of the stress-energy tensor of this effective fluid. This is because it will be coupled to gravitational structures contained in $S_0$ which do satisfy 4D diffeomorphism invariance, and so Bianchi-like identities, implying stress-energy conservation of matter. We showed that this is indeed the case in Section~\ref{Sec:setcons}. In fact, we showed that it could not be otherwise, since the evolution of the Hamiltonian and Momentum dictated by the Dirac algebra, Eqs.~\eqref{dotHH} and \eqref{dotHHi} (inherited by 
matter as \eqref{dotHHm} and \eqref{dotHHim}) are nothing but the Hamiltonian expression of stress-energy conservation. That is the content of the remarkable Eqs.~\eqref{equiv1} and \eqref{equiv2}, the impact of which will only be fully appreciated in the final installment to the series of papers started in~\cite{geoCDM}, where violations of the Dirac algebra will be examined. As shown in~\cite{CDMLIV}, violations of energy conservation are then inevitable. 

We finally included a couple of examples in Section~\ref{examples}, notably identifying $\Sigma_t$ with the Schwarzchild frame. It is tempting to derive phenomenology from the associated accretion phenomena around black holes, but we leave a full treatment to future work.

%\subsubsection{Move to concs}[ARE WE GOING TO HAVE 4 DEGREES OF FREEDOM IN THIS ``FLUID''? YES, IT WOULD SEEM SO, EVEN IF THE ORIGINAL THEORY ONLY HAD AN EXTRA ONE IT MIGHT AS WELL HAVE HAD 4]

%\subsubsection{Shift this to conclusions?}
%[We should ignore where they came from at first, but doesn't this erase phenomenologically interesting correlations? This could be the point...]

%Hence we could even have starting ${\cal H}_i$ and geodesic frames. That would be strange because the preferred frame would not be geodesic before switch off. In a way it is the converse of geodesic to non-geodesic. 

%The most natural would be non-geodesic to non geodesic, but we would have to adapt the early part of the first paper. 

%\subsubsection{checks} It is easy to check, a la Bojowald, that \eqref{Hfluid} and \eqref{Hifluid} satisfy \eqref{smearhihi}-\eqref{smearh0h0}, signs included [but this contradicts Bojo and Thiem signs... check with them]. [ONE LAST PROBLEM WITH \eqref{smearhih0}...] 

%It is difficult to check, but it is plausible, that in a general frame Brown's equations of motion and \eqref{Hfluid}-\eqref{Hifluid} imply the required \eqref{dotHHi} [WITH CORRECTION!!! but this requires using the geodesic equation. [TO BE DONE?]

%\subsubsection{To concs}

%Can we have momentum in a geodesic leaf?

\section{Acknowledgments}
We thank Stephon Alexander, Chris Isham, Shinji Mukohyama and Tom Zlosnik 
%Niayesh Afshordi, Stephon Alexander, Paolo Bassani, Claudia de Rham,  Shinji Mukohyama and Tom Zlosnik 
for help with this paper. 
This work was partly supported by the STFC Consolidated Grants ST/T000791/1 and ST/X00575/1.

\end{document}